\documentclass{jpp}

\usepackage{graphicx}
\usepackage{natbib}

\ifCUPmtlplainloaded \else
  \checkfont{eurm10}
  \iffontfound
    \IfFileExists{upmath.sty}
      {\typeout{^^JFound AMS Euler Roman fonts on the system,
                   using the 'upmath' package.^^J}%
       \usepackage{upmath}}
      {\typeout{^^JFound AMS Euler Roman fonts on the system, but you
                   dont seem to have the}%
       \typeout{'upmath' package installed. JPP.cls can take advantage
                 of these fonts, if you use 'upmath' package.^^J}%
      }
  \else
  \fi
\fi

\ifCUPmtlplainloaded \else
  \checkfont{msam10}
  \iffontfound
    \IfFileExists{amssymb.sty}
      {\typeout{^^JFound AMS Symbol fonts on the system, using the
                'amssymb' package.^^J}%
       \usepackage{amssymb}%

      }{}
  \fi
\fi

\ifCUPmtlplainloaded \else
  \IfFileExists{amsbsy.sty}
    {\typeout{^^JFound the 'amsbsy' package on the system, using it.^^J}%
     \usepackage{amsbsy}}
    {}
\fi

\usepackage{ae}
\usepackage[pdftex]{hyperref} 
	\hypersetup{colorlinks=true,
		pdfstartview=FitV,
		linkcolor=blue,
		citecolor=blue,
		urlcolor=blue,
		pdfauthor={PAPP Gergely, ppg@ipp.mpg.de},
		pdfsubject={2015 Runaway losses in RMP article manuscript},
		pdftitle={2015 Runaway losses in RMP article manuscript}
	}

\newsavebox{\astrutbox}
\sbox{\astrutbox}{\rule[-5pt]{0pt}{20pt}}

\newcommand{\bariota}{\iota\kern-.4em\raisebox{-.05ex}{\large-}}

\title[Energetic electron transport due to magnetic perturbations]{Energetic electron transport in the presence of magnetic
  perturbations in magnetically confined plasmas}

\author[G. Papp, M. Drevlak, G. I. Pokol and T. F\"ul\"op]%
{G.\ns  P\ls a\ls p\ls p$^{1,2}$%
  \thanks{Email address for correspondence: ppg@ipp.mpg.de},\ns
M.\ns D\ls r\ls e\ls v\ls l\ls a\ls k\ls$^2$,
G.\ns I.\ns P\ls o\ls k\ls o\ls l$^3$
\and T.\ns F\ls \"u\ls l\ls \"o\ls p$^4$}

\affiliation{$^1$Max-Planck/Princeton Center for Plasma Physics\\[\affilskip]
$^2$Max-Planck-Institute for Plasma Physics, Garching \& Greifswald, Germany\\[\affilskip]
$^3$Institute of Nuclear Techniques, Budapest University of Technology and Economics, Budapest, Hungary\\[\affilskip]
$^4$Department of Applied Physics, Chalmers University of Technology, Gothenburg, Sweden
}

\pubyear{XXXX}
\volume{XX}
\pagerange{XXX--YYY}
\date{?; revised ?; accepted ?. - To be entered by editorial office}
\begin{document}

\maketitle

\begin{abstract}
The transport of energetic electrons is sensitive to magnetic
  perturbations.  By using 3D numerical simulation of test particle
  drift orbits we show that the transport of untrapped electrons
  through an open region with magnetic perturbations cannot be
  described by a diffusive process. Based on our test particle simulations, we propose a
   model that leads to an exponential loss of particles.
\end{abstract}

\begin{PACS}
41.75.Ht, 52.25.Fi, 52.55.Fa, 52.65.-y
\end{PACS}

\section{Introduction}

There is an increasing
interest in the effect of magnetic perturbations on transport in
fusion plasmas, as it has been shown that externally generated
magnetic perturbations influence many aspects, including e.g. the
presence and properties of Edge Localized Modes \citep{evans04suppression, suttrop11first},
plasma rotation \citep{fietz15influence}, Neoclassical Tearing Modes \citep{koslowski06dependence} or losses of relativistic electrons
\citep{lehnen08suppression}.  Electrons with relativistic velocities
(runaway electrons) are the particles that are most sensitive to magnetic perturbations as
they follow open field lines with $v_\parallel\simeq c$.
The purpose of this work is to provide insights into the effects of magnetic
field perturbations on the transport and loss of energetic electrons by using
three-dimensional (3D) numerical modelling of the electron drift
orbits.\\
\indent Runaway electrons are frequently generated in tokamak disruptions and
can cause damage on plasma facing components. One of the envisaged
mitigation strategies is to use magnetic perturbations to lower the
confinement of the runaway electrons and prevent the formation of a
potentially harmful runaway electron beam. Resonant magnetic
perturbations were tested with some success on various tokamaks
\citep{lehnen08suppression, yoshino99generation,commaux11novel}.
Possible effects and uses of 3D magnetic field perturbations in ITER
have been explored
\citep{shimada07progress,papp11runaway,papp11iter,papp12iter}.  The
individual particle orbits in perturbed magnetic fields are chaotic \citep{papp12iter},
and for a precise calculation of the transport one needs to resort to
3D numerical simulations at large computational expense.  However, for
the purposes of self-consistent modelling of runaway electron
dynamics, it is important to describe the transport due to magnetic
perturbations with a reasonably simple model, so that it can be
included in the modelling of the coupled dynamics of the runaway
electron current and resistive diffusion of the electric field
\citep{papp13effect}. This work aims to provide insight into how such a
simple model can be constructed so that it faithfully reproduces the
dynamics of particle losses due to magnetic perturbations.

\section{Effect of magnetic perturbations} The transport of electrons
in enclosed stochastic magnetic fields is usually described with the
Rechester-Rosenbluth diffusion coefficient $D_{RR} = \pi q v_\| R
\left(\delta B/B\right)^2$~\citep{rechester78electron}. Here $q$ is
the safety factor, $v_\parallel \simeq c$ is the parallel velocity,
$R$ is the major radius {and $\delta B/B\equiv \sqrt{\langle (\delta
    B/B)^2 \rangle_\psi} $ is the flux surface averaged normalized
  magnetic perturbation amplitude as a function of radius.}  In mixed
magnetic topologies consisting of magnetic islands, intact toroidal
magnetic surfaces and stochastic regions, this diffusion coefficient
is not valid \citep{myra92effect,mynick81transport}.  In such
circumstances, some fraction of the electrons will be trapped in
magnetic islands and the confinement time of these particles can equal
the characteristic system evolution time. Furthermore, in open chaotic
systems, such as the edge region of a magnetically confined plasma in
the presence of external magnetic perturbations, not only the
Rechester-Rosenbluth diffusion is not applicable, but the diffusive
approach itself breaks down.  Unlike in the case for enclosed
stochastic fields, the region of increased transport (compared to the
unperturbed case) extends all the way to the wall, making it easier
for particles to get lost.  As was confirmed by numerical simulations
described in this paper, the runaway electron flux starting from a
certain radial position is not proportional to the runaway density
gradient, but the local runaway density itself. If a flat runaway
profile is initialized, losses occur immediately all around the
simulation zone, i.e. the flux is not gradient-driven, as the inital
density gradient is zero. This dynamics is  due to the {\em open} ergodic
region that is created by the RMP coils. Note, that the diffusive
approach may still be applicable in closed ergodic
regions. %Perhaps the simplest
%example of this behaviour is that already at the beginning of the
%system evolution particle losses are observed from regions where the
%initial runaway density was flat. 
The net radial displacement of the
particles is highly sensitive to the initial position and can differ
by several orders of magnitude. Particles born within resonant loss
regions will reach the outer edge of the plasma faster than the rest
and be lost to the wall. Since the system is open at the plasma edge,
there is no return flux of fast electrons, and hence the density
gradient is not the governing factor in the evolution, as it would be
in a diffusive picture.  As we will show, this process can be
approximated with a position dependent exponential loss of particles,
that describes
the ensemble behaviour with reasonable accuracy.\\
\indent In an axisymmetric tokamak equilibrium the magnetic topology
consists of nested magnetic surfaces that provide good particle
confinement. The main loss mechanism for fast electrons is due to the
drift orbit shift associated with the energy gain
\citep{papp11runaway,papp11iter}, which is determined by the electric
field evolution.  In the simulations presented here we use static
externally-imposed magnetic fluctuations on a static equilibrium field
with nested toroidal magnetic surfaces.  Whenever a magnetic
perturbation is present, the topology of the magnetic field is
strongly modified. There appear island chains (the locations of which
are correlated with the low order rational values of the rotational
transform $\bariota = 1/q$), together with intact, but deformed
magnetic surfaces (KAM barriers \cite{kolmogorov54general,
  arnold63small, moser62on}) at irrational $\bariota$ values. In
between these ergodic zones are formed.

We solve the relativistic gyro-averaged equations of motion for
test energetic electrons in the perturbed field with the ANTS code
\citep{papp11runaway}. The simulations have been carried out for a
specific ITER-like scenario as described in \citep{papp11iter}. The
perturbations are externally imposed by the 9x3 quasi-rectangular
perturbation coils at the outer side of the device. In the configuration we use (formerly labeled ``B'' \citep{papp11iter}), electrical
currents flowing in these coils generate a $120^{\circ}$ rotationally
symmetric perturbation field that decays radially towards the inside
of the plasma and results in the relative perturbation strength
$\delta B/B>\mathcal{O}(10^{-3})$ outside the $\psi=0.5$ flux surface (the
normalized radius relates to the normalized flux as $r/a\simeq
\sqrt{\psi}$, where $a$ is the minor radius of the torus).

\begin{figure}
  \begin{center}
\includegraphics[width=0.6\textwidth]{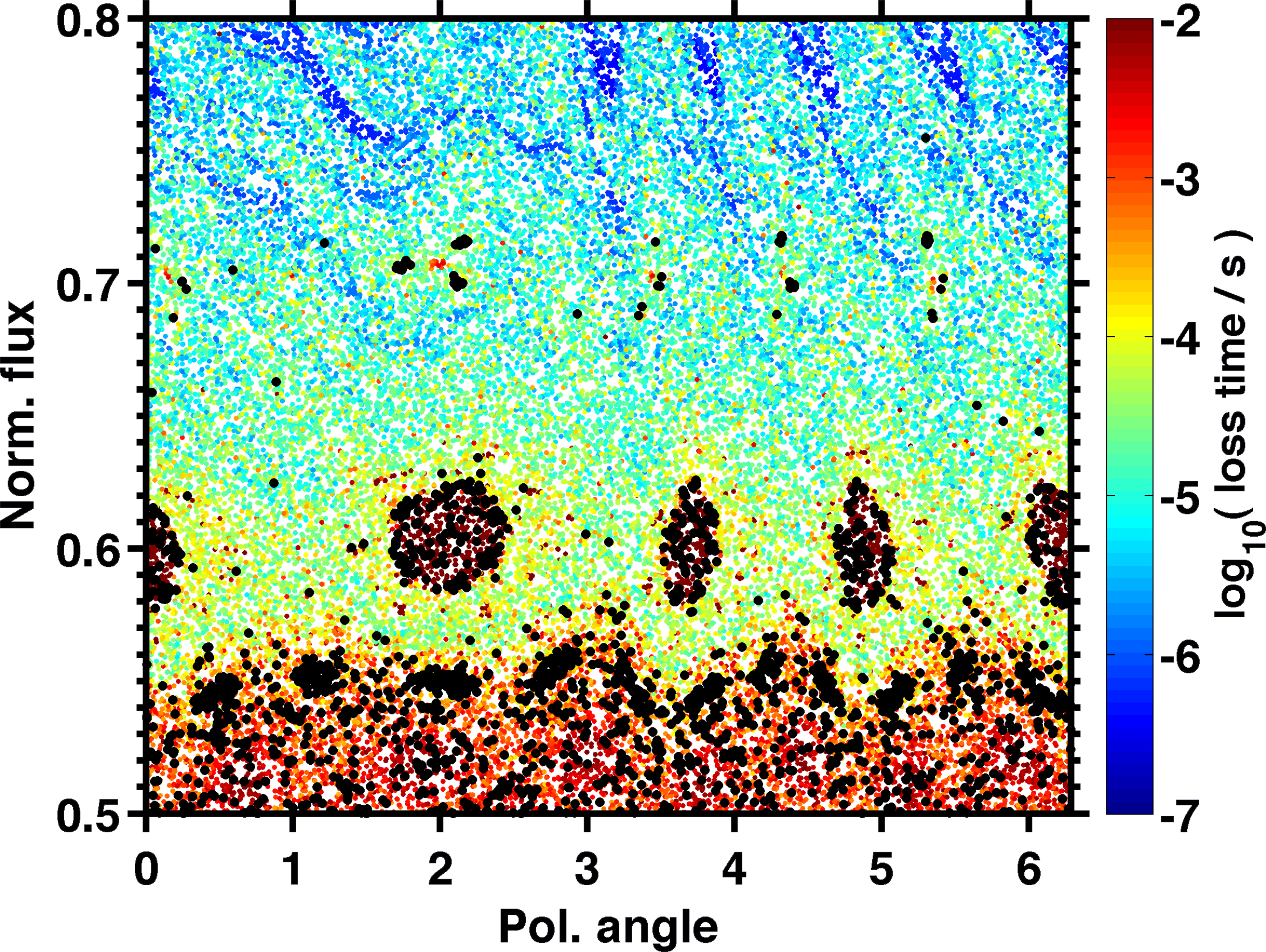}
  \caption{Electron loss time as function of initial coordinates in normalized flux ($\psi$) and poloidal angle ($\vartheta$) at 60 kA coil current.}
\label{fig_poinc}
\end{center}
\end{figure}

To determine the evolution of the particle ensemble in such a topology a direct calculation of
test particle trajectories has been carried out. In each simulation a
minimum of 40~000 test particles with 1 MeV energy were launched parallel to
the magnetic field lines with various starting position
distributions. In the simulations we use a time-dependent electric
field obtained for an ITER-like disruption scenario calculated with a
model of the coupled dynamics of the evolution of the radial profile
of the current density (including the runaways) and the resistive
diffusion of the electric field \citep{smith09runaway}. The electric
field peaks between 6-15~ms, which means that most of the acceleration
happens in this time frame, as it is shown in Fig.~1a of
Ref.~\citep{papp12iter}. The electrons are gaining energy due to the
electric field mainly during this time-interval.
In the unperturbed scenario, the drift orbit shift associated with the energy gain dictates the particle losses, which in this ITER case happens on the $\tau \simeq 10~\mathrm{ms}$ timescale \citep{papp11iter}. 

Figure \ref{fig_poinc} shows the loss time of electrons launched with
($\psi\in$[0.5,0.8],~$\vartheta\in$[0,2$\pi$],~$\phi$=0) as a function of
initial coordinates in the $\phi=0$ poloidal cross section. The
simulation was performed for 100~ms. The particles corresponding to
the black dots have not been lost during this time, as these were
trapped in the remnant O points of island chains formed around
low-order rational surfaces.  Figure \ref{fig_poinc} clearly reveals
the chaotic nature of the process and highlights the extreme
sensitivity of the individual particle loss times to the initial
position.

\begin{figure}
  \begin{center}
\includegraphics[width=0.47\textwidth]{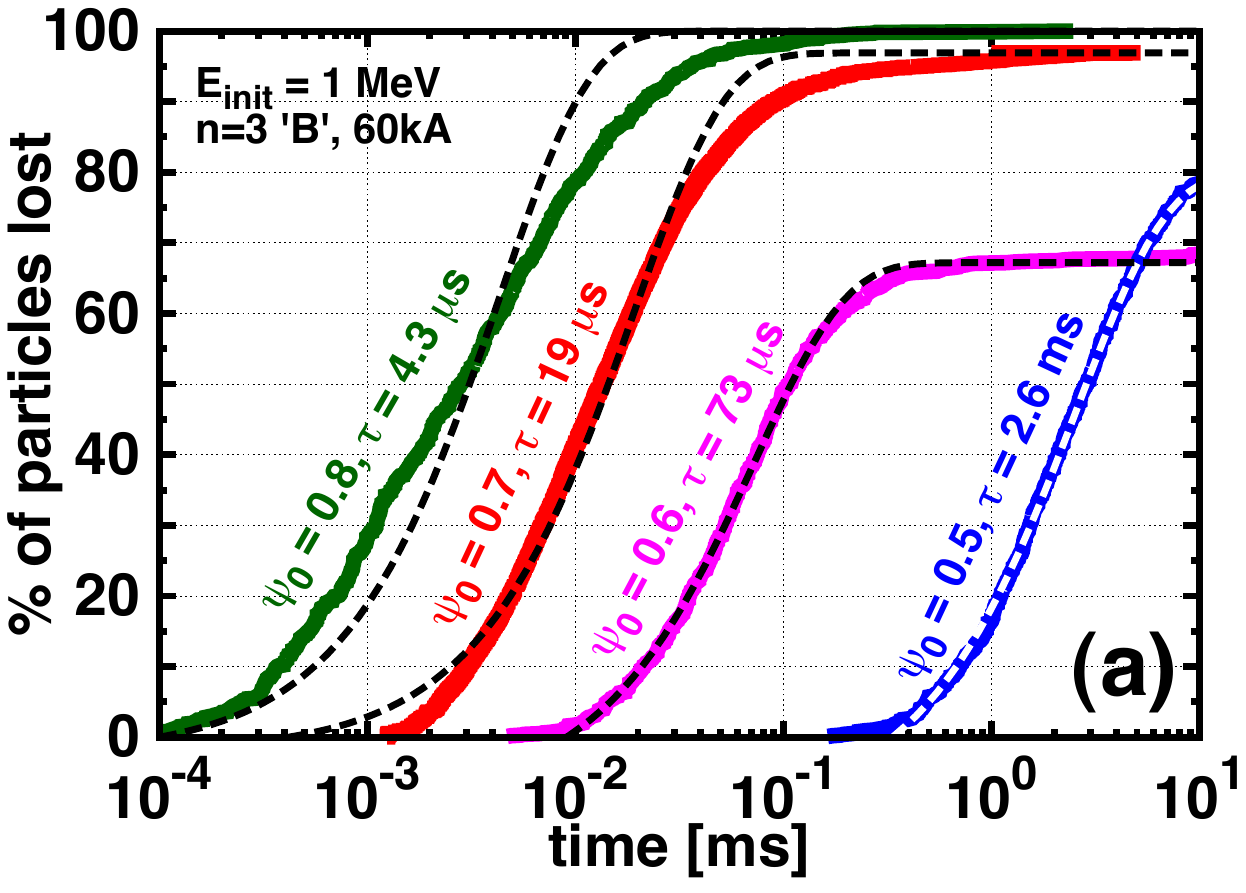}
\includegraphics[width=0.47\textwidth]{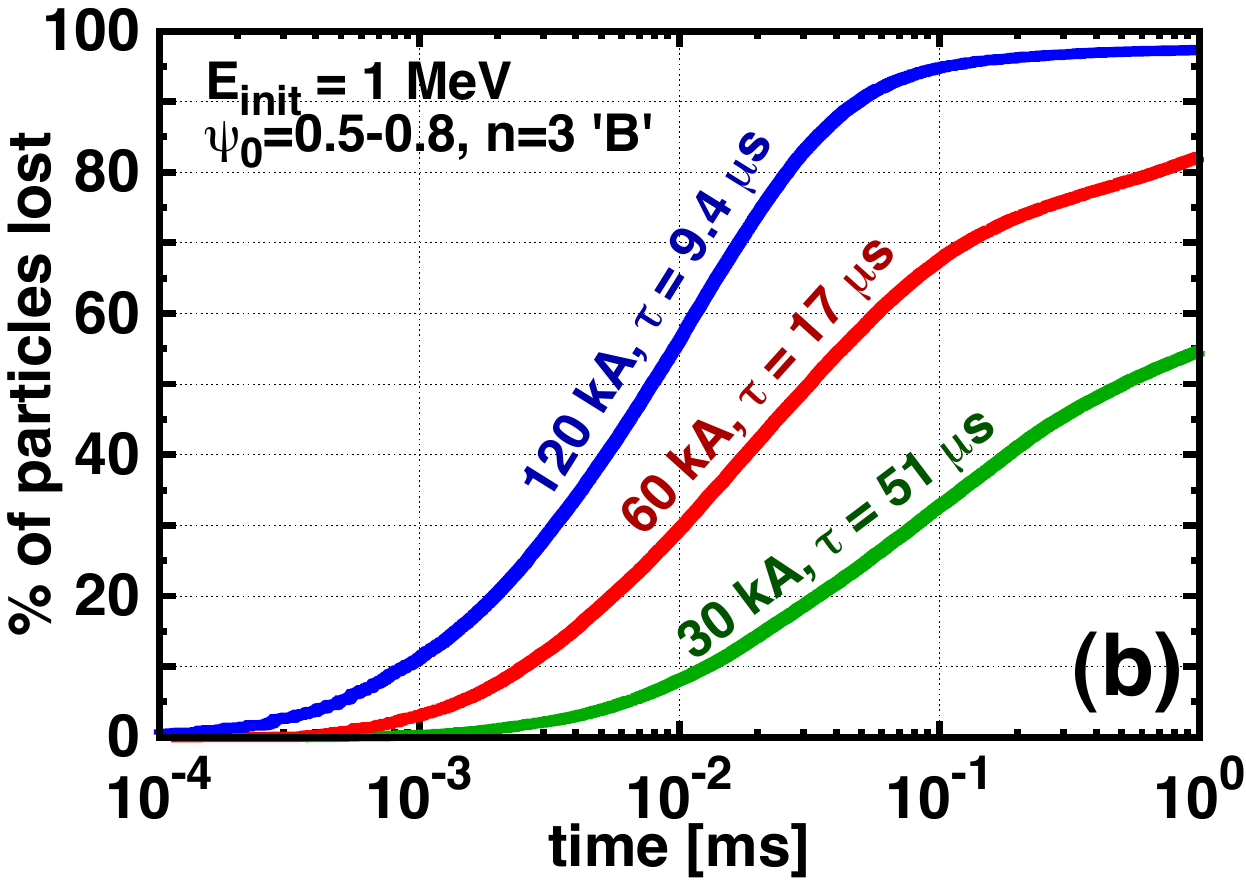}
\caption{(a) Fraction of lost particles for different starting
  radii as a function of time. Dashed lines show $N_\mathrm{lost}(t)= N_0(1- \exp\{-t/\tau\})$ fits.
  (b) Fraction of lost particles within $\psi\in$[0.5,0.8] for different perturbation currents as a function of time.}
\label{fig:loss}
\end{center}
\end{figure}

Despite this complexity, the ensemble behaviour is smooth.
Figure~\ref{fig:loss}a shows the fraction of lost particles as a
function of time for particles launched at individual flux
surfaces. All these curves show an exponential decay of the confined
particles up to the saturation value, determined by the
fraction of particles trapped in remnant O-points of an island chain
intersecting the starting surface. Dashed lines show exponential fits and the characteristic loss times are indicated above the curves.

 The magnetic perturbation enhanced net radial transport leads to
 particle losses on the sub-millisecond timescale. Our energy scans
 confirmed that on this short timescale the behaviour is similar
 regardless of the energy of the particles if $E>1$~MeV, as the particles
 mainly follow the open magnetic field lines with the speed
 $v_\parallel\simeq c$. The long-term ($>10$ ms) behaviour is
 determined by the remnant island structures. Particles starting in
 specific spatial positions can be trapped in the remnant O-points and
 confined even at energies of 100 MeV.

If the magnetic perturbation strength is increased (by increasing the
current in the coils) the island structure will change non-linearly. Therefore the relationship between the time-evolution of the
losses and the magnetic perturbation magnitude is not
trivial. Figure~\ref{fig:loss}b illustrates the fraction of lost
particles as a function of time for three different values of the
perturbation current. To highlight the effect of the magnetic
perturbation magnitude, in these simulations the electric field has
been set to zero (the electrons are not accelerated as they were in
the other simulations).

\begin{figure}
  \begin{center}
\includegraphics[width=0.47\textwidth]{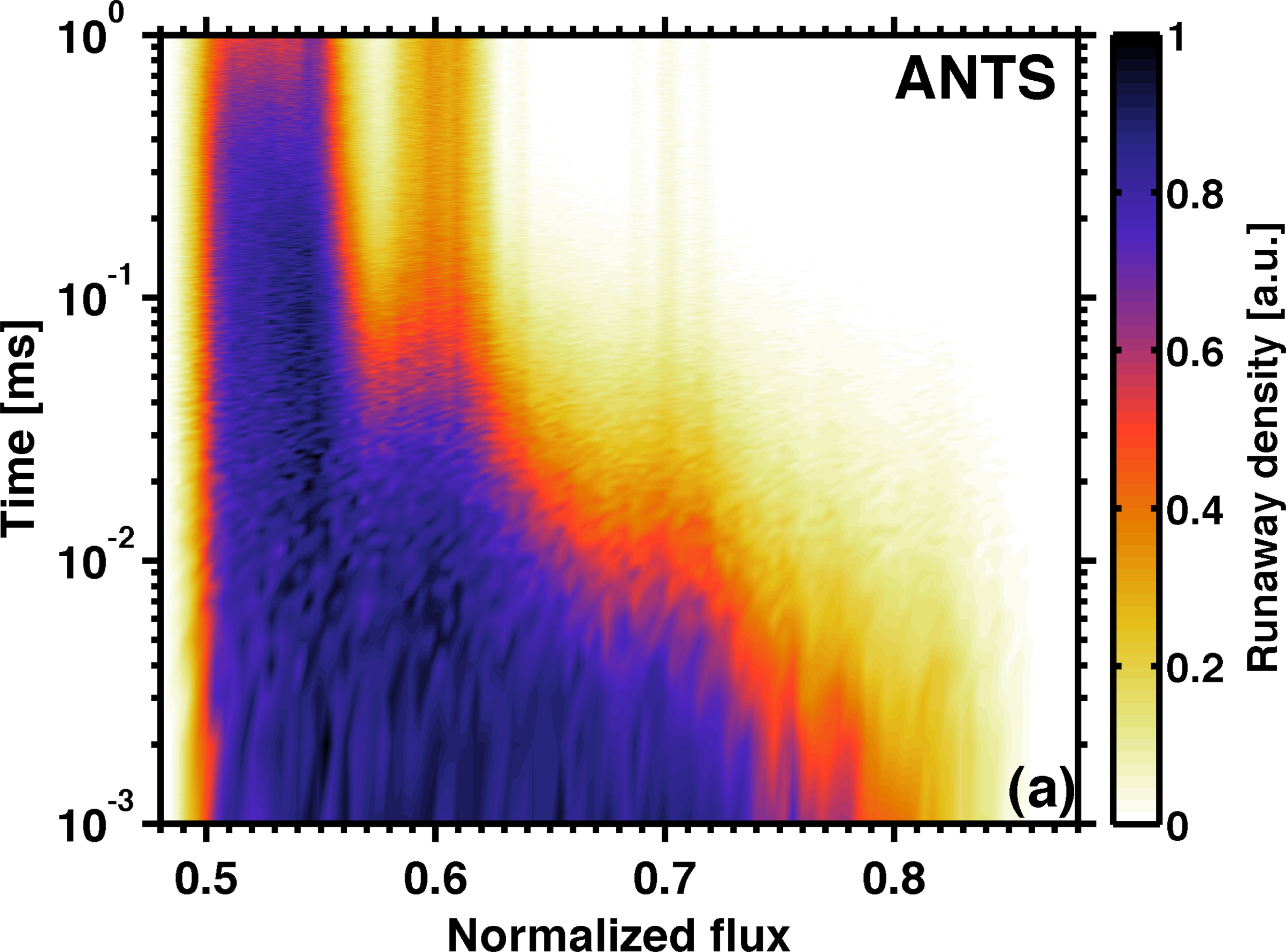}
\includegraphics[width=0.47\textwidth]{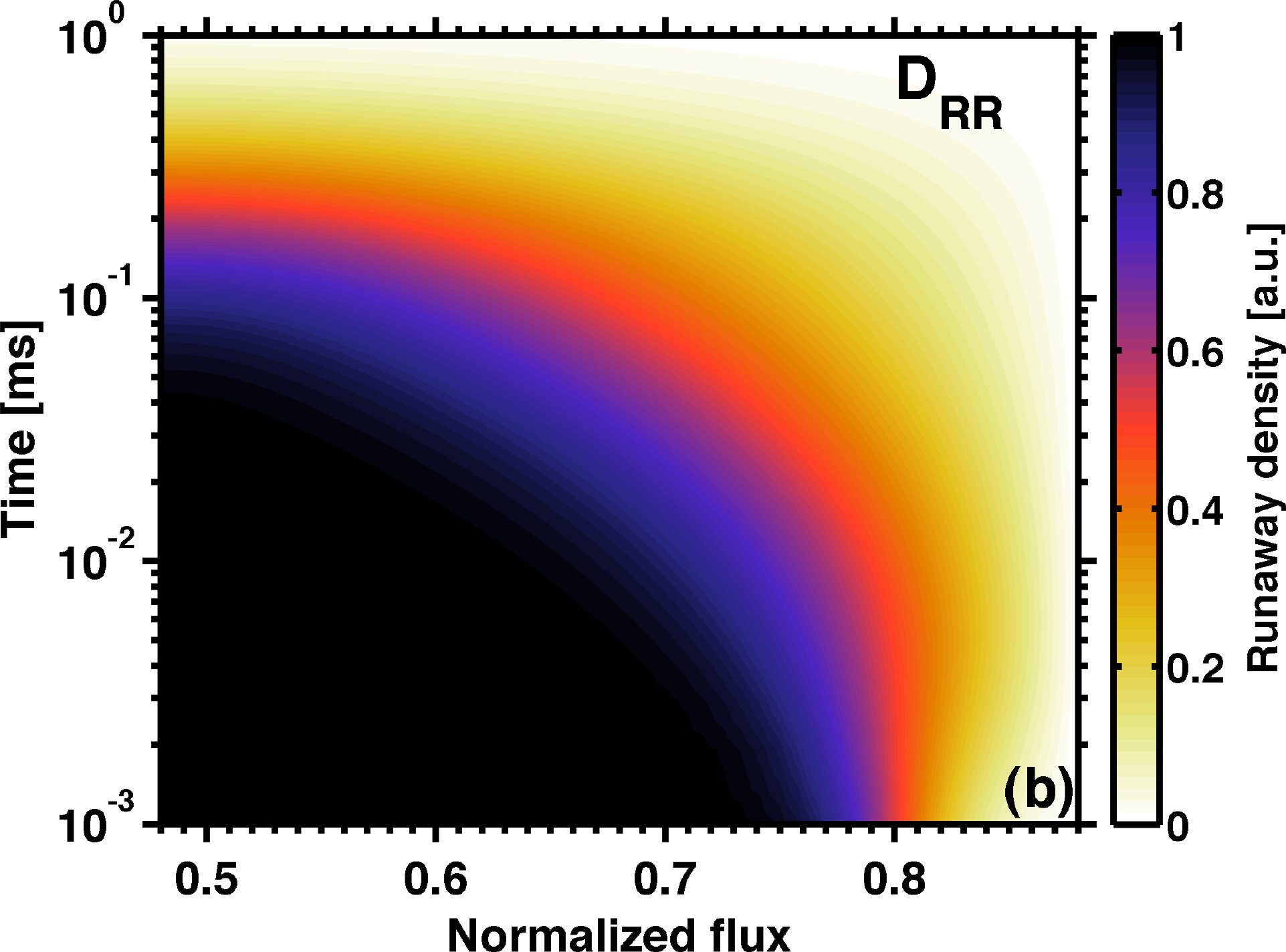}
  \caption{(a) Evolution of fast electron density $n(\psi,t)$ as
    calculated by the ANTS code. \mbox{(b)~Evolution} of the density
    due to Rechester-Rosenbluth diffusion. Both figures are calculated for 60 kA coil current.}
\label{fig:n}
\end{center}
\end{figure}

The smooth ensemble behaviour is also reflected in the evolution of
the electron density as a function of radial position ($\psi$) and time, as is
shown in Fig.~\ref{fig:n}a. The continuous radial loss picture is only
distorted by the enhanced confinement of the remnant island chains at
$\psi=$0.5, 0.6 and 0.7. For comparison, Fig.~\ref{fig:n}b shows the
evolution of density due to the Rechester-Rosenbluth diffusion for the
same initial conditions.  The evolution of the density $n(\psi,t)$ is
largely different: the diffusive model predicts better confinement at the edge but worse confinement in the core, by orders of magnitudes.
The generation of runaway electrons is
exponentially sensitive to the existing amount of runaways through the
avalanche process. If instead of a test particle scenario the runaway
electron generation is included in the calculation,
the differences between the two cases are increased even further, as will be shown
later in this paper.

\begin{figure}
  \begin{center}
\includegraphics[width=0.47\textwidth]{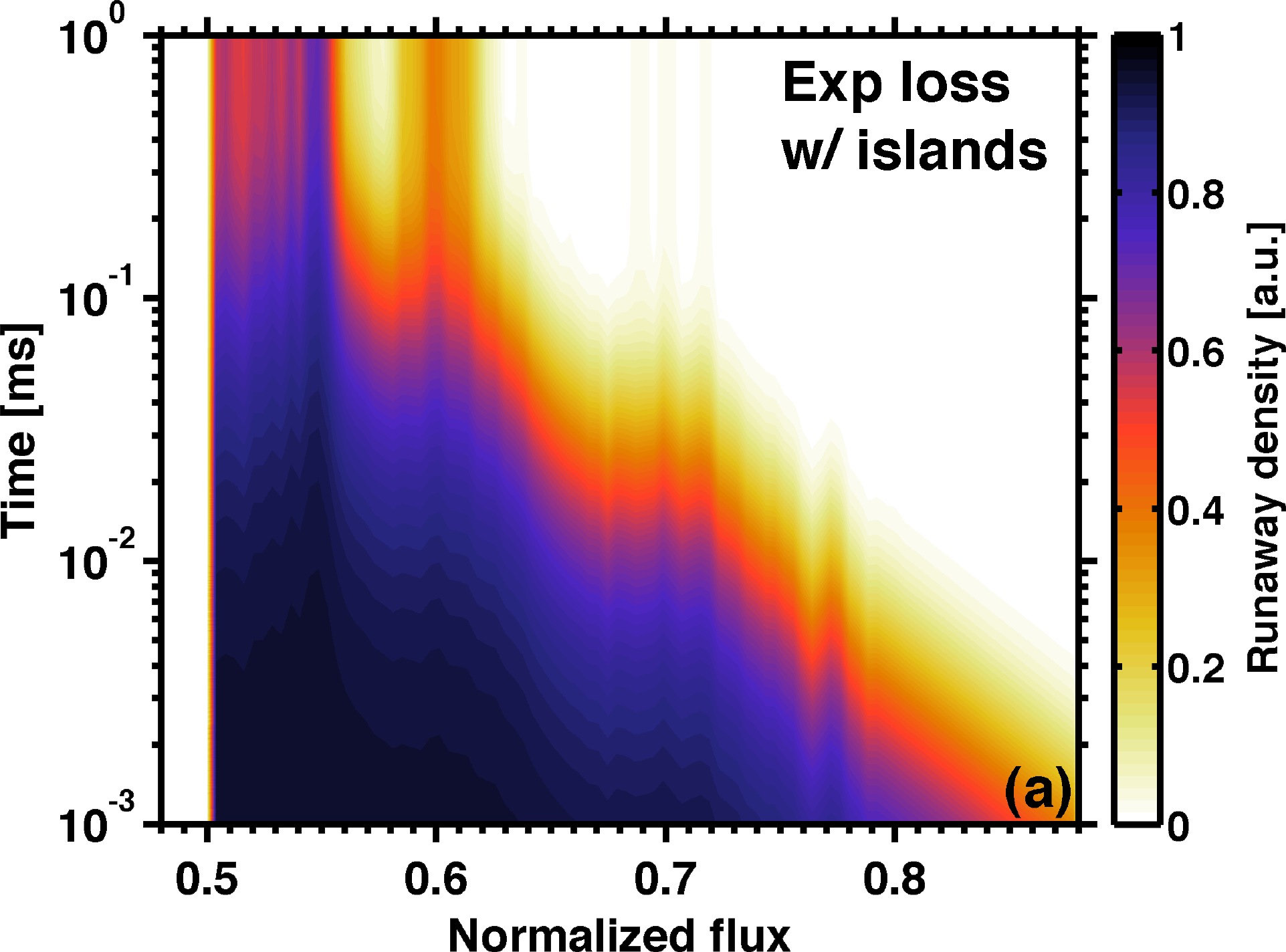}
\includegraphics[width=0.47\textwidth]{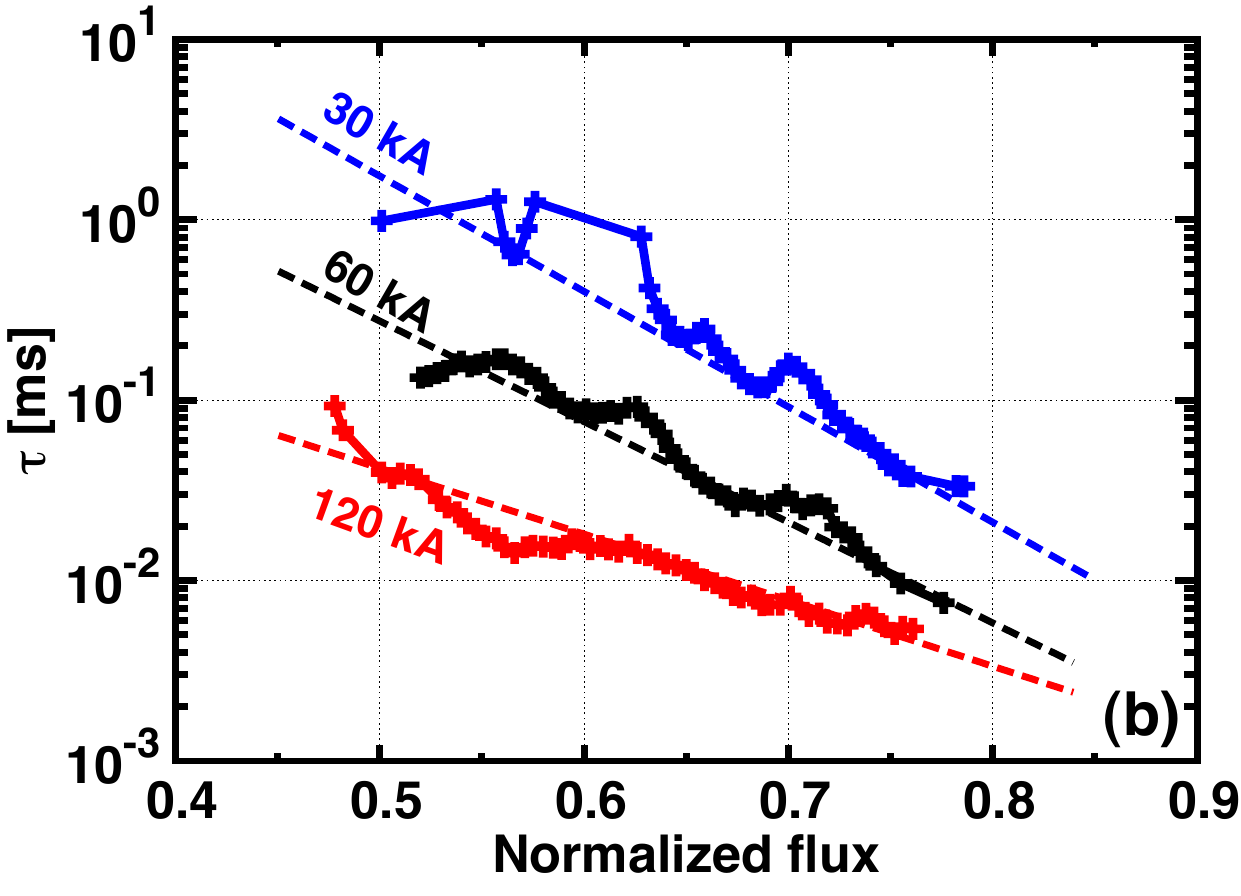}
 \caption{(a) Evolution of electron density using the fitted $\tau(\psi)$ values for the 60 kA configuration, including particles confined in the remnant islands ($n_\mathrm{c}$).
 (b) Fitted values of $\tau$ as a function of normalized flux for 3 different perturbation amplitudes.}
\label{fig:fittedn}
\end{center}
\end{figure}

Following the observations illustrated in figures \ref{fig:loss}a-b
and \ref{fig:n}a, we approximated the particle losses with an
$N_\mathrm{lost}(t)\propto 1- \exp\{-t/\tau(\psi)\}$ trend, where
$\tau(\psi)$ is the characteristic loss time associated with a certain
initial $\psi$ position for the particles. We have performed fits of
$\tau(\psi)$ to the evolution of fast electron density obtained by the
3D test particle simulations for various cases, such as the ones
illustrated in figures \ref{fig:loss}a-b and \ref{fig:n}a. The fitted values of $\tau(\psi)$
are shown in figure \ref{fig:fittedn}b for three different
perturbation magnitudes. We note that the dependence of $\tau(\psi)$
on the radial coordinate is relatively smooth and is well approximated
by an exponential dependence $\tau(\psi)\propto \exp\{-\psi/\psi_0\}$
(dashed lines), where $\psi_0 \propto \delta B / B$. Local maxima in
$\tau(\psi)$ correspond to remnant islands and KAM zones in the
perturbed magnetic topology.

Replacing the diffusive transport in a numerical calculation with the
exponential loss model, we can calculate the evolution of fast
particle density using the fitted $\tau(\psi)$ values. A pure exponential loss description
cannot reproduce the confined fraction. However, if we split the
particle density into a confined and non-confined fraction
$n(\psi,t)=n_\mathrm{c}(\psi)+n_\mathrm{nc}(\psi,t)$, where the confined part is
constant at a given radial position (extracted from the ANTS simulations) and the non-confined part follows the
exponential loss $n_\mathrm{nc}(t,\psi)\propto 1-\exp\{-t/\tau(\psi)\}$, we
can reproduce the results of the 3D simulation. Figure
\ref{fig:fittedn}a shows a calculation based on the exponential losses
for the 60~kA case, including the $n_\mathrm{c}(\psi)$ particles confined in the remnant islands. It clearly resembles the particle evolution
calculated by ANTS (figure \ref{fig:n}a - the figures are placed next
to each other for easier comparison).

\section{Self-consistent calculations} 
\begin{figure}
  \begin{center}
\includegraphics[width=0.50\textwidth]{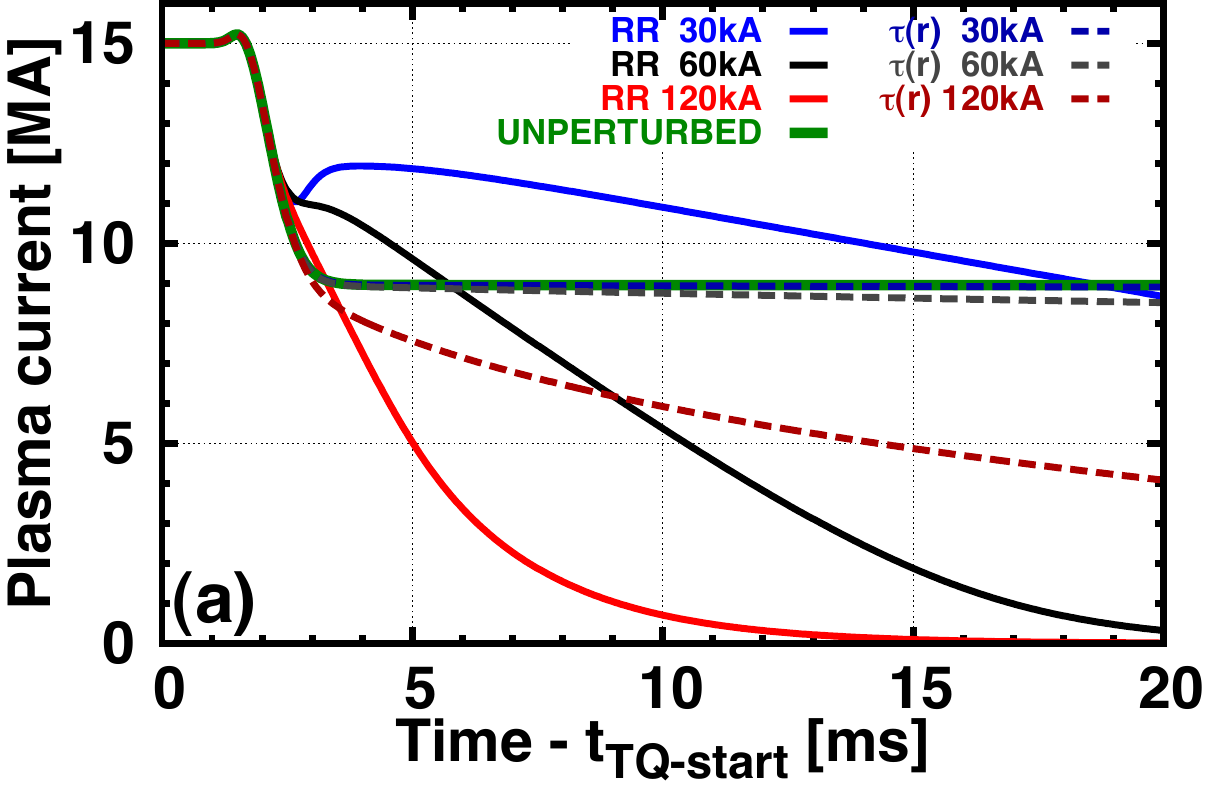}
  \caption{Effect of diffusive (solid lines) vs exponential (dashed lines) losses on plasma (RE) current evolution for ITER in a self-consistent calculation for different perturbation strengths.}
\label{fig:ITER_compare}
\end{center}
\end{figure}

To illustrate the differences between the implications of the
diffusive and exponential-loss models, we have implemented both loss
models in a self-consistent simulation of the runaway current and
electric field evolution \citep{papp13effect}. Figure
\ref{fig:ITER_compare} shows a disruption simulation for the same ITER
scenario that was used for the 3D simulations. The thermal quench
time is set to 0.2~ms and the effective charge $Z_\mathrm{eff}=2$ to
mimic the parameter evolution of massive gas injection simulations
\citep{hollmann15iter,papp13effect}.
As a general conclusion for all cases, following the thermal collapse,
at $\sim$3~ms virtually all of the Ohmic current is replaced by
runaways, and the runaway current consists of $\sim$10\% seed and
$\sim$90\% avalanche current.
Without losses a runaway beam of $\sim$9~MA would form. Using the
Rechester-Rosenbluth diffusion model and the $\delta B / B$ profile
caused by the perturbation coils we get suppression of the runaways
after 20~ms at $I_\mathrm{RMP}=60$~kA and after 10~ms if
$I_\mathrm{RMP}=120$~kA. $I_\mathrm{RMP}=30$~kA not only does not lead
to suppression, but through the redistribution of the primary seed
current leads to temporary stronger avalanching, increasing the overall
runaway current.  However, as we argue in this paper, with the
support of the 3D test particle calculations, the losses in these
systems are not well described by diffusion. Also, the simulations
show that although the perturbation penetrates to the core, the
transport is essentially untouched inside the radius $r/a=0.7$
for $I_\mathrm{RMP}=60$~kA. Therefore, even if the transport is
largely increased towards the edge region in the
exponential loss description, this does not contribute much to
runaway suppression. The reason is that most of the runaways are
formed in the plasma core, where there is good runaway confinement
according to the test particle simulations. As the dashed lines in
figure \ref{fig:ITER_compare} clearly illustrate, {\em virtually no runaway
suppression is achieved} even at $I_\mathrm{RMP}=60$~kA with the exponential loss model. At $I_\mathrm{RMP}=120$~kA we see a slow
drop of the runaway current, however, $I_\mathrm{RMP}>75$~kA
will not be accessible in ITER due to hardware limitations.

\section{Conclusions} The most important conclusion of the present work is
that the transport of the untrapped electrons in open chaotic fields
cannot be described by a diffusive model. Instead there is an
exponential decay of confined particles, where the characteristic loss time is
an exponential function of radius and the perturbation strength. This
difference has a decisive effect on the runaway current and the
associated electric field evolution in a self-consistent
calculation. While diffusive losses would suggest that runaway
avalanches can be avoided by using strong enough externally applied
magnetic perturbations, the  exponential loss model shows that this does not
have to be the case. Although it is difficult to predict exactly the
magnitude of the loss-time and its radial dependence in a specific
scenario without 3D numerical simulations, the exponential dependence
and non-diffusive nature of the transport is robust. Therefore
implementation of such a model in self-consistent fluid simulations should
give important insights into runaway electron dynamics and the prospects of RE mitigation in the presence
of magnetic perturbations.\\

\paragraph{\bf Acknowledgments} The authors are grateful for I.~Pusztai and P.~Helander for fruitful discussions. This research was partially funded by the Max-Planck/Princeton Center for Plasma Physics. This work has been carried out within the framework of the EUROfusion Consortium and has received funding from the European Union's Horizon 2020 research and innovation programme under grant agreement number 633053. The views and opinions expressed herein do not necessarily reflect those of the European Commission. 

\bibliographystyle{jpp}

\bibliography{Papp_pert_JPP2015}

\end{document}